# REVIEW OF SOLID-STATE MODULATORS

E. G. Cook, Lawrence Livermore National Laboratory, USA

*Abstract*

Solid-state modulators for pulsed power applications have been a goal since the first fast high-power semiconductor devices became available. Recent improvements in both the speed and peak power capabilities of semiconductor devices developed for the power conditioning and traction industries have led to a new generation of solid-state switched high power modulators with performance rivaling that of hard tube modulators and thyratron switched line-type modulators. These new solid-state devices offer the promise of higher efficiency and longer lifetimes at a time when availability of standard technologies is becoming questionable.

A brief discussion of circuit topologies and solid-state devices is followed by examples of modulators currently in use or in test. This presentation is intended to give an overview of the current capabilities of solid-state modulators in various applications.

## 1 BACKGROUND

Many of the high-voltage power conditioning requirements of the accelerator community have been satisfied by the use of conventional thyratron, ignitron, sparkgap or, when more waveform control is required, hardtube switches. Modulators using these switching devices have limitations with regard to various combinations of repetition rate, lifetime, efficiency, pulse width agility, average power, cost, and sometimes switch availability.

As accelerator requirements become more demanding, particularly with regard to average power, lifetime, pulsewidth agility, and repetition rate, some of these conventional switching devices are inadequate. However use of solid-state devices in these applications has been held back by device limitations usually in either voltage rating, peak switching power, or switching speed.

## 2 SOLID-STATE DEVICES

The most commonly used fast high-power semiconductor device is the inverter grade thyristor which is available with voltage ratings > 1kV at kA average currents. These thyristors are closing switches that require a current reversal through the device to commutate off, and most high-voltage pulsed-power applications have limited the use of thyristors as replacements for other closing switches. The semiconductor industry's continued development of devices for traction applications and high frequency switching power supplies have created entire families of devices that have a unique combination of switching capabilities. For the first time we are seeing production quantities of devices that may be considered to be close to "ideal switches". These ideal switches are devices that have a fast turn-off capability as well as fast turn-on characteristics; devices that have minimal trigger power requirements and are capable of efficiently switching large amounts of energy in very short periods of time and at high repetition rates if so required.

There are two devices that come close to meeting these criteria for ideal switches, Metal Oxide Semiconductor Field Effect Transistors (MOSFETs) and Insulated Gate Bipolar Transistors (IGBTs). These devices and applications utilizing these devices are the focus of this paper. As shown in Table 1, both MOSFETs and IGBTs are devices that can switch large amounts of power with modest levels of trigger power. MOSFETs have substantially faster switching speeds while IGBTs generally are more efficient, handle more power, and are capable of being manufactured at higher voltage ratings. A very brief explanation of how these devices function follows this section – more detailed information is readily available from vendors and manufacturers.

Table1. MOSFET and IGBT Capabilities

| Parameter | MOSFET | IGBT |
|---|---|---|
| Max. Peak Operating Voltage (V) | 1200 | 3300 |
| Peak Pulsed Current Rating (A) | 100 | 3000 |
| Derated Peak Power (kW/Device)* | >80 | >7000 |
| Switching Speed – ON & OFF (ns) | < 20 | < 200 |
| Gate Controlled Pulsewidth | 20ns-DC | 600ns-DC |
| Control Power (µJ/Pulse) | < 5 | < 30 |
| Device Cost ($/kW switched) | 0.30 | 0.15 |

*MOSFET – Voltage Derating ~ 80% of Max. Vpeak
*IGBT-Voltage Derating ~60% of Max. Vpeak

### 2.1 MOSFETs

As seen in the simplified schematic in Fig. 1, MOSFETs are three terminal devices with the terminals labeled drain, source and gate. MOSFETs are enhancement mode devices that rely upon majority carriers for their operation. Electrically the gate to source connection looks like a capacitor and with zero gate-source voltage the channel (the region between drain and source) is very resistive. As the gate-source voltage is increased, the electric field pulls electrons into the channel and increases the flow of current, i.e. ,the flow of drain-source current is *enhanced* by the gate-source voltage. Once the gate-source capacitance is charged, no additional energy is required to keep the device on.

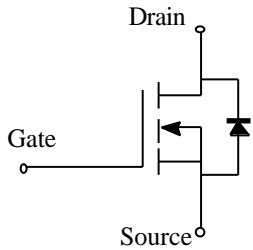

Fig.1 MOSFET Device Symbol (N-type shown)

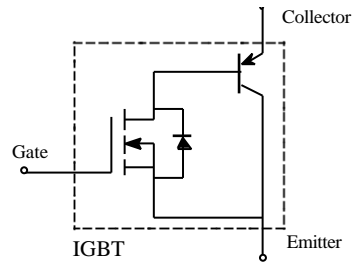

Fig. 2 Simplified Schematic for an IGBT

For pulsed power applications where the goal is to turn the device on very quickly, a fast, large gate current (10's of amperes) is required during turn-on but little power is needed thereafter. Likewise, during turn-off, a large current must be pulled out of the gate-source capacitance. The gate drive circuit must be capable of sourcing and sinking these currents at the required repetition rate.

## *2.2 IGBTs*

As depicted in Fig. 2, an IGBT is also a three-terminal device that combines the high input impedance and gate characteristics of a MOSFET and the low saturation voltage of a bipolar transistor. As the MOSFET is turned on, base current flows in the pnp bipolar transistor, injects carriers into the transistor and turns on the device. While gating off the MOSFET initiates the turn-off process for the transistor, the time required for fast turn-off of the transistor also depends on other factors such as carrier recombination time. Typically both the turn-on and turn-off times for an IGBT are slower than those of a MOSFET, but the peak current density in an IGBT is approximately five times higher than that of a MOSFET having the same die area.

## 3 SWITCH CIRCUIT TOPOLOGY

Single solid-state devices generally don't have the peak voltage rating required for most accelerator applications and, consequently, many devices are usually required for their use in fast high-voltage circuits. Two circuit topologies currently using MOSFETs and IGBTs to achieve these high voltage levels offer significant advantages over other circuit topologies including PFNs, Blumlein lines, and, high step-up ratio transformers. The first circuit approach is to connect as many switching devices in series as is needed to meet the application's requirements. The second approach uses what is commonly referred to as an induction adder where the switching devices drive the primary winding of multiple transformers and the secondary windings of each of the transformers are connected in series to generate the required high voltage. In general, regardless of the switch circuit topology, the total cost of the solid-state devices required to switch a given peak power is determined by the peak power capability of each switch - it matters not whether the devices are arranged in series, parallel, or a combination of both.

When MOSFETs or IGBTs are used, the capabilities of a high-voltage pulsed circuit are greatly enhanced. Since these devices can be gated off as well as gated on, the circuit now has the capability of variable pulsewidth even on a pulse-to-pulse basis, and the circuits may also be operated at high repetition rates. Within the limits of their current rating (which can be increased by paralleling devices), these switches give the circuit topologies a low source impedance, thereby allowing the load to vary over a substantial impedance range. As with all solid-state devices, the expected lifetimes of properly designed circuits are very long.

## *3.1 Series Switch Topology*

A common circuit topology for series stack approach is shown in Fig. 3. In this topology a high voltage power supply charges a DC capacitor bank. The series stack of solid-state devices is connected between the capacitor bank and the load. Gating the stack on and off applies the full bank voltage across the load with the pulsewidth and repetition rate being controlled by the gate trigger pulse. The rise and fall times of the load voltage are determined by the switching characteristics of the specific solid-state devices used in the stack. Implementation of series stack approach requires very careful attention to proper DC and transient voltage grading of the switches in the series stack to force a uniform distribution of voltage across all the series elements under all conditions. All devices must be triggered simultaneously - isolated trigger signals and isolated power sources are usually required. Careful attention to stray capacitance is very important.

Controls for the series stack are also critical, as the control circuits must sense load conditions so faults can be quickly detected and the stack gated off. Stacks assembled with IGBTs can normally sustain short circuit conditions for a short period of time (usually specified to be ~10μs for high power devices) which gives more than adequate time to turn the stack off.

An important operational advantage of the series stack approach is the ability to obtain any desired pulsewidth from the minimum pulsewidth as determined by the capabilities of the switching devices out to and including a DC output. The series circuit topology can yield lower hardware costs but this is partially offset by the additional costs for the voltage grading components and costs associated with achieving the appropriate isolation and/or clearance voltages for the control system, gate circuits, enclosures, etc.

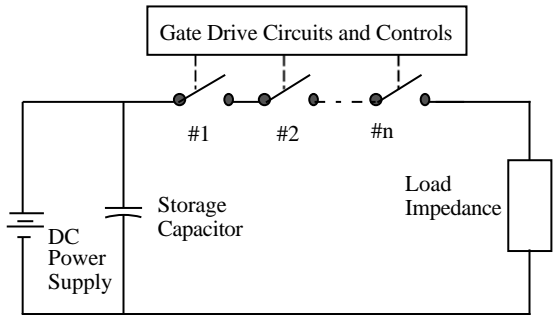

Fig. 3 Typical Series Switch Topology

## 3.2 Adder Topology

In the adder configuration shown in Fig. 4, the secondary windings of a number of 1:1 pulse transformers are connected in series. Typically, for fast pulse applications, both the primary and secondary winding consists of a single turn to keep the leakage inductance small. In this configuration, the voltage on the secondary winding is the sum of all the voltages appearing on the primary windings. An essential criteria is that each primary drive circuit must be able to provide the total secondary current, any additional current loads in the primary circuit, plus the magnetization current for the transformer. This drive current criteria is easily met with the circuit shown in Fig. 4 - the source impedance of a low inductance DC capacitor bank switched by high current IGBTs or a parallel array of MOSFETs or smaller IGBTs is very low (<<1 ). The physical layout for this circuit is important as it is necessary to maintain a small total loop inductance – it doesn't require much inductance to affect performance when the switched di/dt is measured in kA/µs.

In this layout, the solid-state devices are usually ground referenced to take advantage of standard trigger circuits and reduce coupled noise by taking advantage of ground planes. The need for floating and isolated power supplies is also eliminated. The pulse power ground and the drive circuit ground have a common point at the switch source lead but otherwise do not share common current paths thereby reducing switching transients being coupled into the low level gate drive circuits. The lower switch voltages and the corresponding compact conductor loops, i.e., low inductance, enables very fast switching times on the order of tens of nanoseconds.

Voltage grading for individual devices is not a major concern. The transformer provides the isolation between the primary and secondary circuit and if the transformer is designed as a coaxial structure, the high voltage is confined to within the structure. It may help to think of an adder as an induction accelerator with the beam replaced with a conductor.

An adder circuit has a definite maximum pulse width limitation as determined by the available volt-seconds of the transformer magnetic cores. In comparison with the series stack topology, the adder has the additional expense of the transformer mechanical structure including the magnetic cores, and a circuit to reset the magnetic cores between pulses. The gate controls, being ground referenced are simpler and less expensive.

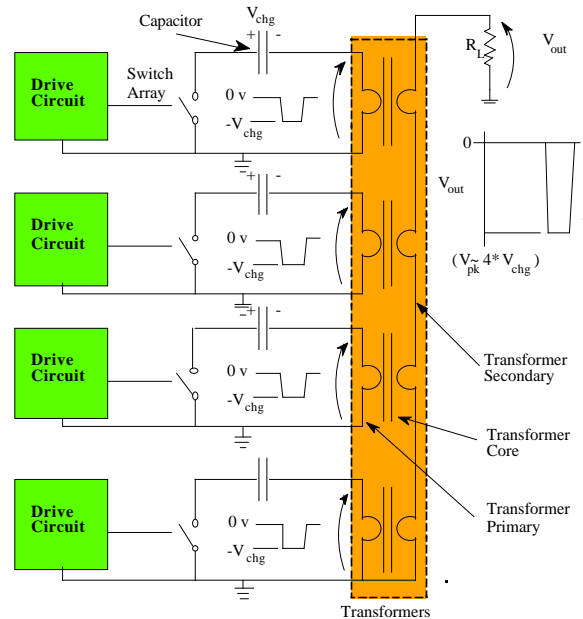

Fig.4 Simplified Schematic of Adder Circuit

## 4 MOSFETS AND IGBTS IN ACCELERATOR APPLICATIONS

Including commercial ventures and government funded projects, MOSFETs and IGBTs are being used in several applications that impact accelerator technology. A few examples are discussed.

### 4.1 Next Linear Collider (NLC) Klystron Modulator

For the past several years, researchers at SLAC have been developing a solid-state modulator based on an IGBT switched adder topology. This aggressive engineering project has made significant progress in demonstrating the capabilities of high-power solid-state modulators. The performance requirements are listed in Table 2.

The IGBTs currently used in the NLC modulator are manufactured by EUPEC and are rated at 3.3kV peak and 800A average current. They have been successfully tested and operated at 3kA peak current at 2.2kV. EUPEC and other manufacturers have devices with higher voltage ratings (>4.5 kV) that are also being evaluated.

Table 2. NLC Klystron Modulator Requirements

| Number of NLC Klystrons | 8 each |
|---|---|
| Operating Pulsed Voltage | 500kV |
| Operating Pulsed Current | 2000 amperes |
| Repetition Rate | 120 Hz. |
| Risetime/Falltime | <200ns 10-90% |
| Flattop Pulse Duration | 3.0µs |
| Energy Efficiency | >75% |

The NLC modulator circuit has demonstrated combined risetime and falltime that meet requirements. A 10-cell prototype capable of generating a 22 kV pulse at 6 kA has been operated as a PFN replacement (~1/10 the volume of the thyrtron switched PFN) to drive a 5045 klystron in the SLAC linac. Fig. 5 is a photo of the 10-cell prototype.

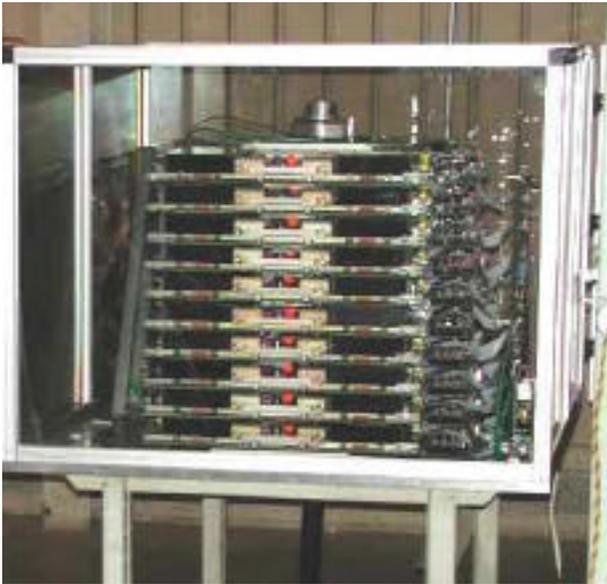

Fig. 5 IGBT Switched Induction Adder - 10 Cell Prototype

### 4.2 Diversified Technologies Inc. (DTI)

DTI manufacturers and markets a broad range of series-switched IGBT modulators that cover a voltage range of up to 150 kV and peak power of 70 megawatts. They have demonstrated switching times of <100ns and pulse repetition rates from DC to 400 kHz. Their hardware has applications in accelerator systems including klystron modulators, ion sources, kicker modulator, and crowbar replacements. These solid-state modulators are specifically designed to compete with and replace vacuum tube based systems. A photograph of one of their PowerMod™ systems is shown in Fig. 6.

### 4.3 Advanced Radiographic Machine (ARM) and Fast Kicker Development at LLNL

The ARM II modulator was one of the first high-power applications of power MOSFETs used in an adder configuration. Its specifications are listed in Table 3. A photo of a single adder module is shown in Fig. 7.

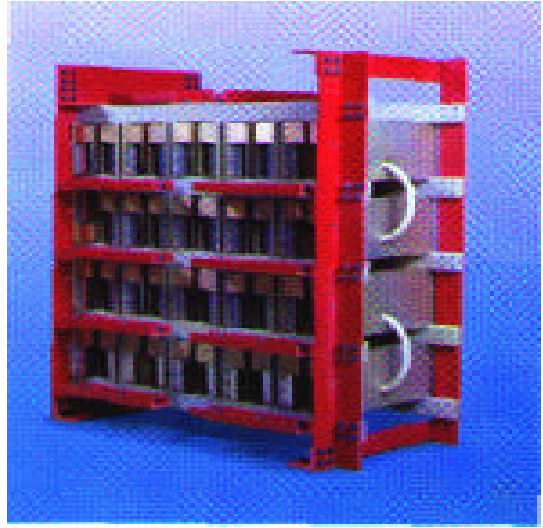

Fig. 6 DTI's 125kV, 400A Solid-State Switch

Table 3. ARM II Modulator Specifications

| Design Voltage | 45 kV (15kV/Adder Module) |
|---|---|
| Maximum Current | 4.8-6 kA |
| Pulsewidth | 200ns-1.5µs |
| Maximum Burst PRF | 1 Mhz |
| Number of MOSFETs | 4032 |

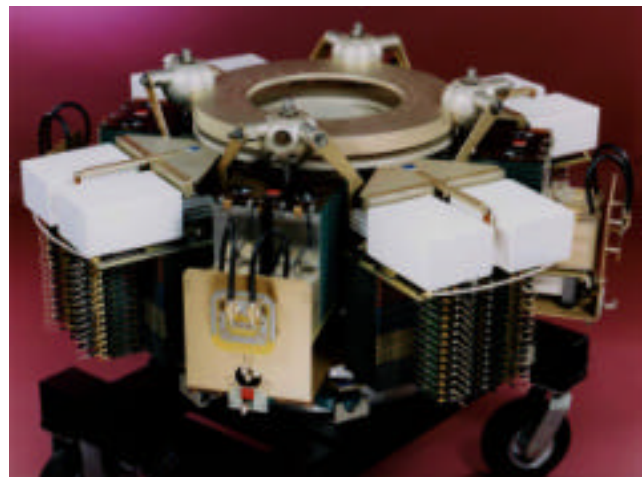

Fig.7 ARM II Module

Fast kicker pulser development at LLNL has been based on the previous ARM modulator development. As listed in Table 4, the major differences in requirements are related to the faster risetime, falltime, and minimum pulsewidth. To control the stray inductance, each parallel array of MOSFETs drives a single pulse transformer. A photo of the kicker assembly is shown in Fig. 8.

Table 4. Fast Kicker Specifications

| Output Voltage | 20 kV into 50 |
|---|---|
| Voltage Risetime/Falltime | <10 ns 10-90% |
| Pulsewidth | 16ns-200ns variable within burst |
| Burst Frequency | >1.6 MHz - 4 pulses |

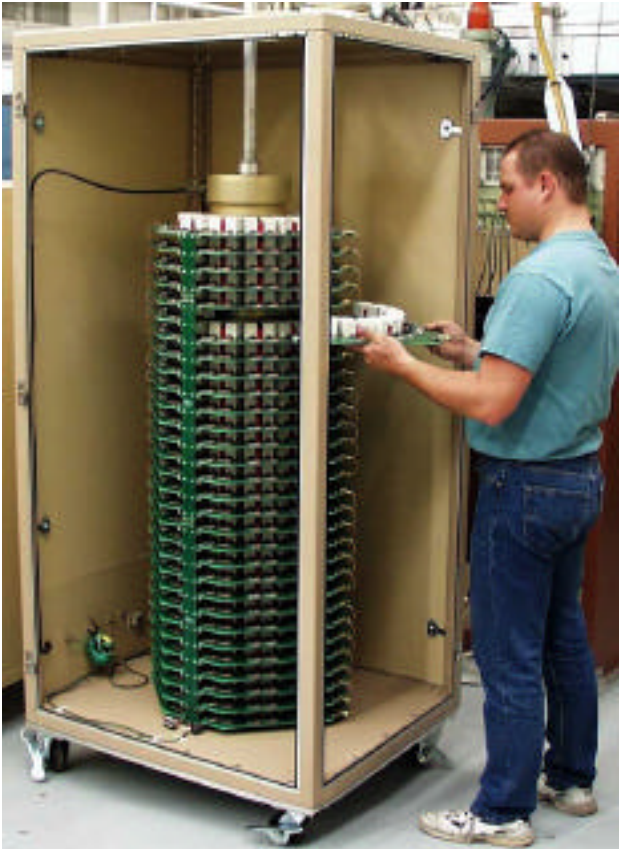

Fig. 8 Photo of Complete Fast Kicker Assembly

*4.4 First Point Scientific*

First Point Scientific has developed a variety of MOSFET switched adder systems. Based on their Miniature Induction Adder (MIA - see Fig. 9), they have demonstrated high repetition rate, high-voltage systems for air pollution control. First Point Scientific has also developed a prototype for an economical, fast, high repetition rate modulator featuring pulse width agility and waveform control for a small recirculator to be used in ion accelerator experiments.

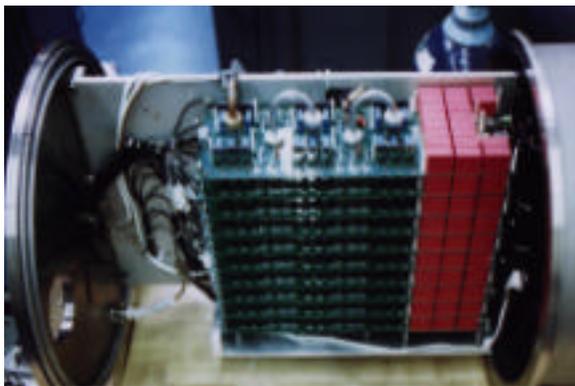

Fig. 9 First Point Scientific - Miniature Induction Adder

## 5 CONCLUSIONS

Faster and higher power solid-state devices are constantly being introduced that offer significant advantages for pulse power applications. These devices are being incorporated into a significant number of modulator designs and used in various projects for specific accelerator applications. As the performance of these devices continues to improve, they will replace more of the conventional switch technologies.

## REFERENCES

[1] H. Kirbie, et al, "MHz Repetition Rate Solid-State Driver for High Current Induction Accelerators", *1999 Part. Accel. Conf.*, New York City, Mar.29-April 2, 1999, http:ftp.pac99.bnl.gov/Papers/
[2] R. Cassel, "Solid State Induction Modulator Replacement for the Conventional SLAC 5045 Klystrons Modulator", *LINAC 2000 - XX International Linac Conf.*, Monterey, CA, August 21-25, 2000
[3] W. J. DeHope, et al, "Recent Advances in Kicker Pulser Technology for Linear Induction Accelerators", *12$^{th}$ IEEE Intl. Pulsed Power Conf.*, Monterey, CA, June 27-30, 1999
[4] Yong-Ho Chung, Craig P. Burkhart, et al, "All Solid-state Switched Pulser for Air Pollution Control System", *12$^{th}$ IEEE Intl. Pulsed Power Conf.*, Monterey, CA, June 27-30, 1999
[5] M. Gaudreau, et al, "Solid State Modulators for Klystron/Gyrotron Conditioning, Testing, and Operation", *12$^{th}$ IEEE Intl. Pulsed Power Conf.*, Monterey, CA, June 27-30, 1999
[6] E. Cook, et al, "Inductive Adder Kicker Modulator for DARHT-2", *LINAC 2000 - XX International Linac Conf.*, Monterey, CA, August 21-25, 2000